\documentclass[journal,twocolumn]{IEEEtran}
\ifCLASSINFOpdf
\else
\fi
\usepackage{graphicx}          % Include this line if your
\usepackage{ifpdf}
\usepackage{cite}
\usepackage[numbers,sort&compress]{natbib}
\usepackage{amsmath}
\usepackage{amsthm}
\usepackage{epsfig}
\usepackage{color}
\usepackage{enumerate}
\usepackage{epstopdf}

\begin{document}
\title{Adaptive Approximation-Based Control for Nonlinear Systems: A Unified Solution with Accurate and Inaccurate Measurements}

%\author{Michael~Shell,~\IEEEmembership{Member,~IEEE,}
%        John~Doe,~\IEEEmembership{Fellow,~OSA,}
%        and~Jane~Doe,~\IEEEmembership{Life~Fellow,~IEEE}% <-this % stops a space
%\thanks{M. Shell was with the Department
%of Electrical and Computer Engineering, Georgia Institute of Technology, Atlanta,

\author{Dong Zhao
%\author{Michael~Shell,~\IEEEmembership{Member,~IEEE,}
%        John~Doe,~\IEEEmembership{Fellow,~OSA,}
%        and~Jane~Doe,~\IEEEmembership{Life~Fellow,~IEEE}% <-this % stops a space
%\thanks{This work was supported by the Beijing Natural Science Foundation
%under Grant 4232047 and the National Natural Science Foundation of
%China under Grant 62273021.%in part by the Shandong Provincial
%%Natural Science Foundation, China, under Grant ZR2017QF007, in part by the
%%National Natural Science Foundation of China under Grant 61233014, and Grant
%%61773242, and in part by the National High Technology
%%Research and Development Program of China under Grant 2015AA042201. (Corresponding author: Yueyang Li)
%}
\thanks{Dong Zhao is with the School of Cyber Science and Technology, Beihang
University, Beijing 100191, China (e-mail: dzhao@buaa.edu.cn).}% <-this % stops a space

\thanks{Manuscript received XX XX, XX; revised XX, XX.}}


\markboth{Journal of \LaTeX\ Class Files,~Vol.~xx, No.~x, x~2020}%
{Shell \MakeLowercase{\textit{et al.}}: Bare Demo of IEEEtran.cls for IEEE Journals}



\maketitle

\begin{abstract}
A unified solution to adaptive approximation-based control for nonlinear systems with accurate and inaccurate state measurement is synthesized in this study. Starting from the standard adaptive approximation-based controller with accurate state measurement, its corresponding physical interpretation, stability conclusion, and learning ability are rigorously addressed when facing additive measurement inaccuracy, and explicit answers are obtained in the framework of both controller matching and system matching. Finally, it proves that, with a certain condition, the standard adaptive approximation-based controller works as a unified solution for the cases with accurate and inaccurate measurement, and the solution can be extended to the nonlinear system control problems with extra unknown dynamics or faults in actuator and/or process dynamics. A single-link robot arm example is used for the simulation demonstration of the unified solution.
\end{abstract}


% \fi
\IEEEpeerreviewmaketitle

\section{Introduction}

%\IEEEPARstart{W}{ith} the
%sdf
Adaptive approximation-based control is one of the most important and influential method for control systems \cite{polybook}. Based on the functional approximation technique and the adaptive control technique, adaptive approximation-based control can accomplish control tasks with partial or fully unknown plant dynamics. Motivated by its significant application potential, adaptive approximation-based control has attracted extensive research interest since it was proposed. The major results aim at the complex nonlinear system control, where neural network or fuzzy approximation is commonly used \cite{ge2007,tong2017,ka2015}. The detailed design for different system descriptions has been well studied, e.g., switched systems \cite{switched}, large-scale systems \cite{pana2008}, stochastic systems \cite{sto}, and multi-agent systems \cite{hua2017}. Moreover, the adaptive approximation-based control owns a wide application for solving fault issues \cite{zhang2004,yin2017}.

For the control system design, accurate measurement is vital, since it provides the feedback of current system status and directs the next control move. For approximation-based control, the requirement of accurate measurement is more intensive: measurement feedback contributes directly to approximate the unknown system dynamics (effect), and then the control task can be done with unknown system dynamics. Under this circumstance, accurate measurement is commonly assumed for adaptive approximation-based control \cite{polybook,ge2007,tong2017,ka2015,switched,pana2008,sto,hua2017,zhang2004,yin2017}. However, the situation in practice is not as perfect as assumed that accurate measurement is not always available, where measurement noise, disturbance, sensor fault, and/or cyber attack are not inevitable. It should be noted that these inaccuracies in measurement cannot always be identified due to their intrinsic stealthiness, and accordingly the controller reconfiguration is unavailable. Even if the process monitoring or fault detection is employed online, the inaccurate measurement will last at least for some time without being noticed. Besides, due to the unknown or uncertain process dynamics, pure model-based measurement reconstruction is unsatisfactory, especially for nonlinear systems.

Since controller reconfiguration is not always available regarding the occurrence of inaccurate measurement, for the adaptive approximation-based controller driven by accurate measurement, one question is natural: does this controller work either when facing inaccurate measurement? Furthermore, it is essential to know: how well and on which condition does this controller work? Again, since inaccurate measurement will cause a mismatch in the closed-loop control system, the physical interpretation or problem reformulation is needed. Finally, although different adaptive approximation-based controllers have been proposed for dealing with different scenarios \cite{songsb,jin2017,yu2020,khe2015,khe2016,khe2018,bou2018}, including process fault, (partially) unknown process dynamics, and actuator faults, the unified solution has not been well addressed with performance analysis and existence conditions.

To give a deep insight into these questions and provide the answers, in this study, the properties of the standard adaptive approximation-based controller with both accurate and inaccurate measurements for general nonlinear uncertain systems are analyzed. For the same controller with different inputs, we establish two frameworks of problem formulation: indirect stabilization and direct stabilization. The control performance with inaccurate measurement and the standard adaptive approximation-based controller is explicitly presented, accompanied by new assumptions and physical interpretations. It shows that the adaptive approximation-based controller designed with accurate measurement works as a unified solution for the inaccurate measurement case as well. If one knows the existence of additive measurement inaccuracy, the ``active" adaptive approximation-based controller design is given as in \cite{songsb,jin2017,yu2020,khe2015,khe2016,khe2018,bou2018}; however, it is covered by the unified solution as well (detailed discussions are given in Section IV). Furthermore, proved by the framework of indirect stabilization and direct stabilization, the general problems, e.g., extra unknown actuator dynamics or process dynamics for the nonlinear control systems, can be covered by the standard adaptive approximation-based controller designed as well. The major contribution of this study is to present a new viewpoint for adaptive approximation-based control with accurate and inaccurate measurement, and reveals the conditions, performance, and unified solution to various control problems with the adaptive approximation-based controller.

The rest of this paper is organized as follows. The adaptive approximation-based control with accurate measurement is given in Section II, and the controller analysis with inaccurate measurement is given in Section III. In Section IV, generalization discussions regarding the controller stability, required assumptions, and unified solution are provided. In Section V, simulation study is presented, and Section VI concludes this paper.

\textbf{Notation}. $P>0$ denotes the positive definite matrix $P$. For vector $x$, $\left \| x \right \|$ denotes its Euclidean norm. For the functions in this study, we omit the time variable $t$ or state variable $x$ for simplicity if it is clear from the context.
%I wish you the best of success.

\section{Adaptive Approximation Based Control with Accurate Measurement}
In this section, the adaptive approximation-based control solution with accurate state measurement for a general nonlinear uncertain system is given, including the explicit controller design, related assumptions, and control performance analysis.
\subsection{System Description}
Consider the following nonlinear uncertain systems:
\begin{eqnarray}\label{1}
\dot x = f(x) + gu + \eta (x,t) + h(x,t),
\end{eqnarray}
where $x \in R^n$ is the system state and $u \in R^m$ is the system input; $f(x): R^n \mapsto R^n$ and $g \in R^{n \times m}$ denote the nominal dynamics of the system; $\eta (x,t): R^n \times R^+ \mapsto R^n$ is the partially known modeling uncertainty; and $h(x,t): R^n \times R^+ \mapsto R^n$ are the fully unknown dynamics or system fault vector.

The system described in \eqref{1} covers a large range of nonlinear systems, including fully known dynamics ($f$ and $g$), partially known dynamics ($\eta$), and fully unknown dynamics ($h$). The dynamics denoted by $\eta$ is optional, which is modeled only when its necessary information, e.g., upper bound or bounding function, is known and can be used for control design; otherwise, $\eta$ is lumped into $h$. The partially known dynamics is introduced here to keep the generality of the model and also the availability of model information for a targeted controller design by using the model information thoroughly.

\subsection{Controller Design with Accurate Measurement}
To focus on the controller design, we consider a full state measurement and feedback. The following controller is then designed:
\begin{eqnarray}\label{2}
u(t) = {u_0}(x) + {u_c}(x),
\end{eqnarray}
where $u_0$ is the \emph{nominal control effort} regarding the known and partially known system dynamics, and $u_c$ is the \emph{compensation control effort} regarding the generalized unknown system dynamics in the control system (the generalized unknown dynamics implies not only $h$, and more cases will be further analyzed later). The modularization of control effort depends on both the model and control system information acquiring.

\subsubsection{Nominal Control Effort Design}

For $u_0$, the following inequality is expected:
\begin{eqnarray}\label{3}
\frac{{\partial V(x)}}{{\partial x}}\{ f(x) + g{u_0}(x) + \eta (x,t)\}  \le  - \alpha (\left\| x \right\|),
\end{eqnarray}
where
\begin{eqnarray}\label{4}
{\alpha _1}(\left\| x \right\|) \le V(x) \le {\alpha _2}(\left\| x \right\|),
\end{eqnarray}
$\alpha _1$, $\alpha _2$, and $\alpha$ are all ${\kappa _\infty }$ functions \cite{HKbook}, and $V(x)$ is a Lyapunov function for system \eqref{1}. Based on \eqref{3} and \eqref{4}, it is known that $u_0$ can stabilize system \eqref{1} when $h=0$. To better motivate the nominal control effort design, the following example is presented.

\emph{Example 1. Consider a detailed system description for system \eqref{1}:
\begin{eqnarray}\label{5}
\dot x = Ax + B\left[ {f(x) + gu + \eta (x,t)} \right],
\end{eqnarray}
where
\begin{displaymath}
A = \left[ {\begin{array}{*{20}{c}}
{{{\bf{0}}_{(n - 1) \times 1}}}&{{I_{n - 1}}}\\
0&{{{\bf{0}}_{1 \times (n - 1)}}}
\end{array}} \right],~~B = \left[ {\begin{array}{*{20}{c}}
{{0_{n - 1}}}\\
1
\end{array}} \right],
\end{displaymath}
$g \ne 0$, $\left\| {\eta (x,t)} \right\| \le \bar \eta (x,t)$, and $\bar \eta (x,t)$ is the known bounding function. Let
\begin{eqnarray}\label{7}
u(t) = {u_0}(x) = \frac{{ - Kx - {\mathop{\rm sgn}} ({B^T}Px)\bar \eta (x,t)}}{g},
\end{eqnarray}
where $A - BK$ is Hurwitz, and for any $Q>0$, ${(A - BK)^T}P + P(A - BK) =  - Q$ has a solution $P > 0$. Furthermore, if $V = \frac{1}{2}{x^T}Px$, we have $\alpha (\left\| x \right\|) = \frac{1}{2}{x^T}Qx$ according to \eqref{3}.}
\subsubsection{Compensation Control Effort Design}
It is a general task to design $u(t)$ for \eqref{1} with $h=0$, while only $u_0$ is needed, especially when system \eqref{1} is feedback linearizable. For the compensation control effort, due to the lack of information of $h$, the adaptive approximation-based control design technique is used.

When $h \ne 0$, following \eqref{3}, we have
\begin{eqnarray}\label{8}
\dot V \le  - \alpha (\left\| x \right\|) + \frac{{\partial V(x)}}{{\partial x}}\{ g{u_c}(x) + h(x,t)\} .
\end{eqnarray}
To keep the control performance, comparing \eqref{8} with \eqref{3}, it requires that
\begin{eqnarray}\label{9}
\frac{{\partial V(x)}}{{\partial x}}\{ g{u_c}(x) + h(x,t)\}  \le \gamma.
\end{eqnarray}
where $\gamma\ge 0$ is an unknown constant.

Since general nonlinear uncertain systems are considered, the inequality in \eqref{9} characterizes the expected control performance when facing unknown system dynamics, which implies certain relationship between the structure of the unknown dynamics and the system input \cite{HKbook}, e.g., the direct matching condition between system input and unknown dynamics, or lower triangular system structure allowing a backstepping design). Similar assumptions have been introduced in the literature for fault accommodation design.

In general, it is difficult to ensure the same level of control performance for system \eqref{1} with and without facing unknown system dynamics. Thus, $\gamma$ is introduced to model the potential control performance degradation due to the unknown system dynamics. For simplicity and aiming at the approximation-based control, a constant bound $\gamma$ is adopted, which can be replaced by a bounding function with respect to the system state without changing the major results of this paper.

For the inequality in \eqref{9} with respect to $u_c$, the solution is non-unique. Define the control effort $u_c$ corresponding to the minimum of $\gamma$ as $u_c^*$ such that
\begin{eqnarray}\label{10}
\frac{{\partial V(x)}}{{\partial x}}\{  - gu_c^* + h(x,t)\}  \le \gamma^*.
\end{eqnarray}
where $\gamma ^*$ is the minimum of $\gamma$.

Directly, it is known that $u_c^*$ is driven by state $x$, i.e., $u_c^*(x)$. Since $u_c^*(x)$ is not structured, the adaptive approximation for $u_c^*(x)$ as $\hat u_c (x)$ is introduced. For implementation, the following linear approximation network is employed:
\begin{eqnarray}\label{11}
{\hat u_c}(x,\vartheta ) = \Pi (x)\vartheta ,
\end{eqnarray}
where $\Pi (x) = diag\{ \pi _1^T(x), \cdots ,\pi _m^T(x)\} $ is the known basis matrix, $\pi_l: {R^n} \mapsto {R^{n_u^l}}$ is the basis function vector ($l \in \{ 1,...,m\} $), and $\vartheta  \in {R^{{\bar n_u}}}$ is the weight vector ($\bar n_u=\sum\limits_{l=1}^{m}{n_u^l}$). To measure the performance of the approximation, the optimal weight vector $\theta $ is introduced:
\begin{eqnarray}\label{12}
\theta  = \arg \left\{ {\mathop {\inf }\limits_{\vartheta  \in \Theta } \left\{ {\mathop {\sup }\limits_{x \in X} \left\| {u_c^*(x) - {{\hat u}_c}(x,\vartheta )} \right\|} \right\}} \right\},
\end{eqnarray}
where $\Theta  $ is the parameter space for the weight vector and $X \subset R^n$. Based on $\theta$, the gap between $u_c^*(x)$ and $\hat u_c (x,\vartheta)$ corresponds to the following assumption.

\textbf{Assumption 1.} For any $x \in X$, it holds that $\left\| {u_c^*(x) - {{\hat u}_c}(x,\theta )} \right\| \le \lambda $, where $\lambda  \in {R^ + }$ is the unknown bounding parameter.

Since $\theta$ and $\lambda$ are all unknown, their estimations, $\hat \theta$ and $\hat \lambda$, are introduced for a feasible control effort implementation, respectively.

Thus, the adaptive approximation-based compensation control effort is designed as
%\begin{align}\label{12}
%{u_c}(x) = & - \Pi (x)\hat \theta  - \hat \lambda s(x),\\
%\dot {\hat \theta}  =&~ \Gamma {\Pi ^T}(x){p^T}(x) - \Gamma {w_\theta }\hat \theta ,\\
%\dot {\hat \lambda}  =&~ \Lambda p(x)s(x) - {w_\lambda }\hat \lambda ,
%\end{align}
\begin{eqnarray}\label{13}
{u_c}(x) =  - \Pi (x)\hat \theta  - \hat \lambda s(x),
\end{eqnarray}
\begin{eqnarray}\label{14}
\dot {\hat \theta}  = \Gamma {\Pi ^T}(x){p^T}(x) - \Gamma {w_\theta }\hat \theta ,
\end{eqnarray}
\begin{eqnarray}\label{15}
\dot {\hat \lambda}  = \Lambda p(x)s(x) - {w_\lambda }\hat \lambda ,
\end{eqnarray}
where
\begin{displaymath}
p(x) = \frac{{\partial V(x)}}{{\partial x}}g,~~s(x) = \tanh \left( {\frac{{{p^T}(x)}}{\omega }} \right),
\end{displaymath}
$\Gamma >0$ and $\Lambda >0$ are the learning rates for $\hat \theta$ and $\hat \lambda$, respectively, and ${w_\theta } \ge 0$, ${w_\lambda } \ge 0$, and $\omega  > 0$ are the controller parameters.

To check the stability of the closed-loop system, the controller dynamics are included. Thus, the Lyapunov function for the closed-loop system is augmented to be
\begin{eqnarray}\label{16}
{V_c} = V(x) + \frac{1}{2}{\tilde \theta ^T}{\Gamma ^{ - 1}}\tilde \theta  + \frac{1}{{2\Lambda }}{\tilde \lambda ^2},
\end{eqnarray}
where $V(x)$ satisfies \eqref{3} and \eqref{4}, $\tilde \theta  = \hat \theta  - \theta $, and $\tilde \lambda  = \hat \lambda  - \lambda $. Based on \eqref{8} and \eqref{10}, the time derivative of $V_c$ can be deduced as
\begin{align}\label{17}
{{\dot V}_c} \le & - \alpha (\left\| x \right\|) + \frac{{\partial V(x)}}{{\partial x}}\{ g{u_c}(x)\nonumber\\
 &+ h(x,t)\}  + {\dot {\hat \theta} ^T}{\Gamma ^{ - 1}}\tilde \theta  + {\Lambda ^{ - 1}}\tilde \lambda \dot {\hat \lambda} \nonumber\\
 \le & - \alpha (\left\| x \right\|) + p(x)\{ u_c^*(x) + {u_c}(x)\} \nonumber\\
 &+ {\dot {\hat \theta} ^T}{\Gamma ^{ - 1}}\tilde \theta  + {\Lambda ^{ - 1}}\tilde \lambda \dot {\hat \lambda} +\gamma^*\nonumber\\
 \le & - \alpha (\left\| x \right\|) + p(x)\{ u_c^*(x) - {{\hat u}_c}(x,\theta )+ {{\hat u}_c}(x,\theta )\nonumber\\
 & + {u_c}(x)\}  + {\dot {\hat \theta} ^T}{\Gamma ^{ - 1}}\tilde \theta  + {\Lambda ^{ - 1}}\tilde \lambda \dot {\hat \lambda}+\gamma^*.
\end{align}
Based on Assumption 1 and taking \eqref{13}-\eqref{15} into \eqref{17}, it yields
\begin{align}\label{18}
{{\dot V}_c} \le & - \alpha (\left\| x \right\|) + \left\| {p(x)} \right\|\lambda  + p(x)\nonumber\\
 &\times \{ \Pi (x)(\hat \theta  - \tilde \theta ) - \Pi (x)\hat \theta  - \hat \lambda s(x)\} \nonumber\\
 &+ {\dot {\hat \theta }^T}{\Gamma ^{ - 1}}\tilde \theta  + {\Lambda ^{ - 1}}\tilde \lambda \dot {\hat \lambda} +\gamma^*\nonumber\\
 \le & - \alpha (\left\| x \right\|) + \left\| {p(x)} \right\|\lambda  - \hat \lambda p(x)s(x)\nonumber\\
& - {w_\theta }{{\hat \theta }^T}\tilde \theta  - \frac{{{w_\lambda }\tilde \lambda \hat \lambda }}{\Lambda } + \tilde \lambda p(x)s(x)+\gamma^*\nonumber\\
 \le &  - \alpha (\left\| x \right\|) + \left\| {p(x)} \right\|\lambda  - {w_\theta }{{\hat \theta }^T}\tilde \theta \nonumber\\
& - \frac{{{w_\lambda }\tilde \lambda \hat \lambda }}{\Lambda } - \lambda p(x)s(x)+\gamma^*.
\end{align}
The following Lemma will be used for a further derivation.

\textbf{Lemma 1.} For any $a \in R^n$ and $b>0$, it holds that
\begin{displaymath}
\left\| a \right\| - {a^T}\tanh (\frac{a}{b}) \le nbc,
\end{displaymath}
where $c$ is the solution of $c = {e^{ - (c + 1)}}$ ($c \approx 0.2785$).

\textbf{Proof}. Denote $a_i$ as the $i$-th element of vector $a$, where $i \in \{ 1,...,n\} $. Based on the fact that $\left| {{a_i}} \right| - {a_i}\tanh (\frac{{{a_i}}}{b}) \le bc$ \cite{polyauto}, it comes that
\begin{displaymath}
\begin{array}{l}
\left\| a \right\| - {a^T}\tanh (\frac{a}{b}) \le \sum\limits_{i = 1}^n {\left| {{a_i}} \right|}  - \sum\limits_{i = 1}^n {{a_i}\tanh (\frac{{{a_i}}}{b})} \le nbc.
\end{array}
\end{displaymath}
It completes the proof.

Based on Lemma 1 and completing the squares for the cross terms in \eqref{18}, we have
\begin{eqnarray}\label{19}
{\dot V_c} \le  - \alpha (\left\| x \right\|) - \frac{{{w_\theta }}}{2}{\tilde \theta ^T}\tilde \theta  - \frac{{{w_\lambda }}}{{2\Lambda }}{\tilde \lambda ^2} + \varepsilon ,
\end{eqnarray}
where $\varepsilon  = \frac{{{w_\theta }}}{2}{\theta ^T}\theta  + \frac{{{w_\lambda }}}{{2\Lambda }}{\lambda ^2} + \lambda m\omega c+\gamma^*$. Based on \eqref{19}, it can be known that the system state and the adaptive parameter estimation errors are uniformly ultimately bounded \cite{dzhao2022,HKbook}. Furthermore, one can set i) ${w_\theta }$, ${w_\lambda }$, and $\omega $ to be sufficiently small to ensure an arbitrary small final bound for state and these parameter estimation errors or ii) ${w_\theta } = {w_\lambda } = 0$ and $s(x) = {\mathop{\rm sgn}} \left( {{p^T}(x)} \right)$ to ensure $\varepsilon \rightarrow 0$ and then $\mathop {\lim }\limits_{t \to \infty } \left\| {x(t)} \right\| \to 0$.

The performance regarding the controller in \eqref{2} with \eqref{13} is summarized below.

\textbf{Theorem 1.} Under Assumption 1, the controller in \eqref{2} with $u_0 (x)$ satisfying \eqref{3} and $u_c(x)$ given by \eqref{13} guarantees that the state of system \eqref{1} is uniformly ultimately bounded.

The following example is given to enrich the details of the compensation control effort design.

\emph{Example 2. Following Example 1, we consider the following system:
\begin{eqnarray}\label{20}
\dot x = Ax + B\left[ {f(x) + gu + \eta (x,t) + h(x)} \right].
\end{eqnarray}
Based on \eqref{10}, it is known that $u_c^*(x) = \frac{{h(x)}}{g}$. For the compensation control effort design, it is known that $p(x) = {x^T}PBg$, and the rest of $u_c$ is the same as given in \eqref{13}-\eqref{15}.}

\textbf{Remark 1.} Two special cases for the compensation control effort design can be found in the literature. Case 1: if the matching condition is satisfied (e.g., Example 2), the estimation of $h$ is straightforward and thus \eqref{10} holds, e.g., \cite{poly2001}. Case 2: if all the system dynamics are unknown, the approximator is adopted to approximate both $u_0$ and $u_c^*$, e.g., \cite{liu2015}. That is to say that we can find a direct scheme based on control effort compensation or an indirect scheme based on unknown dynamics compensation. The choice of the direct or indirect scheme depends on the sufficiency of model information, e.g., the mentioned matching condition and system dynamics. Sufficient model information can help to direct the approximation and improve the control performance (e.g., $\gamma^*=0$ when the matching condition is known to be satisfied between system input and the unknown dynamics). Thus a trade-off between the approximation burden and prior model information is needed. To keep the generality, here we use the model information for the design of $u_0$ and also consider the approximation directly for the optimal compensation control effort, which covers both the mentioned cases.

\textbf{Remark 2.} In this study, constant bounds $\gamma$ and $\lambda$ are introduced to characterize the corresponding performance limits due to the unknown system dynamics and approximation error, respectively. In addition to these constant bounds, the major results of this study can be extended to the case with state-related bounding functions directly. To focus mainly on the unified solution of the approximation-based control, the constant bounds are adopted.
\section{Adaptive Approximation Based Control with Inaccurate Measurement}
In Section II, the controller design is based on accurate measurement, and a satisfying control performance can be achieved easily. Due to the disturbance, noise, fault, cyber attack, or event-induced measurement error, the measurement may not be accurate in practice. To model the inaccurate measurement, following Section II, we have
\begin{eqnarray}\label{21}
y(t) = x(t) + s(t),
\end{eqnarray}
where $y(t)$ is the accessible measurement for the controller and $s(t) \in R^n$ is the inaccurate portion in $y(t)$. To focus on the control system analysis, here we employ the additive form inaccuracy model and the multiplicative case will be discussed later.

With the inaccurate measurement as in \eqref{21}, the controller in \eqref{2} turns to be
\begin{eqnarray}\label{22}
u(t) = {u_0}(y) + {u_c}(y),
\end{eqnarray}
where $u_0 (y)$ and $u_c(y)$ are obtained by replacing $x$ in $u_0(x)$ satisfying \eqref{3} and $u_c(y)$ in \eqref{13}  by $y$, respectively. Thus, Inheriting from the properties of $u_0(x)$, it holds that
\begin{eqnarray}\label{23}
\frac{{\partial V(y)}}{{\partial y}}\{ f(y) + g{u_0}(y) + \eta (y,t)\}  \le  - \alpha (\left\| y \right\|),
\end{eqnarray}
where
\begin{displaymath}
{\alpha _1}(\left\| y \right\|) \le V(y) \le {\alpha _2}(\left\| y \right\|).
\end{displaymath}

Regarding the inaccurate measurement and the shift of the controller from \eqref{2} to \eqref{22}, the following questions are to be answered:
\begin{description}
  \item[i)] How to understand the controller given by \eqref{22}?
  \item[ii)] How is the control performance of the control system with the controller given by \eqref{22}?
  \item[iii)] In which condition can the controller given by \eqref{22} guarantee the (bounded) control performance for system \eqref{1}?
\end{description}
Question i) is about the physical interpretation of the controller in \eqref{22}, e.g., which part is approximated or compensated when \eqref{21} is activated? Since the measurement is changed, the explanation of the controller may change as well for direct or indirect scheme (Remark 1). Due to the imperfection of model information and multiple-sourced inaccurate signal, it may not be easy to use the observer-based technique to reconstruct the accurate measurement efficiently (this is further discussed in Section IV). Most importantly, measurement inaccuracy has its intrinsic stealthiness. Even if the monitoring or detection technique is used, the conservativeness of the detection threshold or model inaccuracy results in the unawareness of the measurement inaccuracy. Based on these facts, without controller reconfiguration, the controller given by \eqref{22} will be kept but driven by inaccurate feedback signal, and thus the problems ii) and iii) are essential.

In the following, we will analyze the control system for (1) with (21) and (22) from two aspects: indirect stabilization and direct stabilization. For indirect stabilization, the analysis is based on the stabilization of the measurement $y$, and the direct stabilization analysis is based on system state $x$.

\subsection{Indirect Stabilization}
Rewrite the control system with respect to the inaccurate measurement $y$ to match the controller in \eqref{22}. Based on \eqref{1} and \eqref{21}, we have
\begin{eqnarray}\label{24}
\dot y = f(y) + gu + \eta (y,t) + {h_y}(y,t),
\end{eqnarray}
where
\begin{displaymath}
\begin{array}{l}
\begin{aligned}
{h_y}(y,t) = &~\dot s(t) + f(x)  + \eta (x,t)+ h(x,t)\\
 &- f(y)  - \eta (y,t)-h_y(y,t).
\end{aligned}
\end{array}
\end{displaymath}
Here, it can be seen that $h_y(y,t)$ is the lumped dynamics due to the model mismatch with respect to the inaccurate measurement $y$. Based on \eqref{22}-\eqref{24}, it yields
\begin{align}\label{25}
\dot V(y) = &\frac{{\partial V(y)}}{{\partial y}}\{ f(y) + g{u_0}(y) + \eta (y,t)\} \nonumber\\
 &+ \frac{{\partial V(y)}}{{\partial y}}\{ g{u_c}(y) + {h_y}(y,t)\} \nonumber\\
 \le &  - \alpha (\left\| y \right\|) + \frac{{\partial V(y)}}{{\partial y}}\{ g{u_c}(y) + {h_y}(y,t)\} .
\end{align}
As $u$ is driven by $y$, an optimal compensation control effort $u_{c,y}^*(y)$ that corresponds to the non-negative bound $\gamma_y^*$ is expected to satisfy that
\begin{eqnarray}\label{26}
\frac{{\partial V(y)}}{{\partial y}}\{  - gu_{c,y}^*(y) + {h_y}(y,t)\}  \le \gamma_y^*.
\end{eqnarray}
Following the derivation procedure for the adaptive approximation-based controller for $u_c(x)$ in \eqref{13}, the optimal approximation for $u_{c,y}^*(y)$ is denoted as ${\hat u_{c,y}}(y,{\theta _y})$, which is obtained by replacing $x$ and $\theta$ in ${\hat u_c}(x,\theta )$ by $y$ and $\theta_y$, respectively, and
\begin{displaymath}
{\theta _y} = \arg \left\{ {\mathop {\inf }\limits_{\vartheta  \in {\Theta }} \left\{ {\mathop {\sup }\limits_{y \in Y} \left\| {u_{c,y}^*(y) - {{\hat u}_{c,y}}(y,\vartheta )} \right\|} \right\}} \right\},
\end{displaymath}
where $\theta _y $ is the optimal weight vector and $Y \subset {R^n}$. To derive the conclusion for ${u_c}(y)$, the following assumption is made by following Assumption 1.

\textbf{Assumption 2.} For $y \in Y$, it holds that $\left\| {u_{c,y}^*(y) - {{\hat u}_{c,y}}(y,{\theta _y})} \right\| \le {\lambda _y}$, where ${\lambda _y} \in {R^ + }$ is the unknown bounding parameter.

Let ${V_{c,y}} = V(y) + \frac{1}{2}{\tilde \theta_y ^T}{\Gamma ^{ - 1}}\tilde \theta_y  + \frac{1}{{2\Lambda }}{\tilde \lambda_y ^2}$, where ${\tilde \theta _y} = \hat \theta  - {\theta _y}$ and ${\tilde \lambda _y} = \hat \lambda  - {\lambda _y}$. Note that, now $\hat \theta$ and $\hat \lambda$ for the adaptive approximation-based controller are driven by $y$ as mentioned for $u_c(y)$, and they are aimed as the estimations for $\theta _y$ and $\lambda _y$, respectively. Based on Assumption 2, controller \eqref{22}, system \eqref{24}, and the inequality in \eqref{25}, after some manipulations, it yields
\begin{eqnarray}\label{28}
{\dot V_{c,y}} \le  - \alpha (\left\| y \right\|) - \frac{{{w_\theta }}}{2}\tilde \theta _y^T{\tilde \theta _y} - \frac{{{w_\lambda }}}{{2\Lambda }}\tilde \lambda _y^2 + {\varepsilon _y},
\end{eqnarray}
where ${\varepsilon _y} = \frac{{{w_\theta }}}{2}\theta _y^T{\theta _y} + \frac{{{w_\lambda }}}{{2\Lambda }}\lambda _y^2 + \lambda m\omega c+\gamma_y^*$.

Based on \eqref{28} and Theorem 1, it is known that the control performance for system \eqref{1} with \eqref{21} and \eqref{22} will be $\left\| y \right\| \to 0$ (when ${\varepsilon _y} \to 0$ ). Thus, for system \eqref{1} with \eqref{21}, it can be known that
\begin{description}
  \item[Ii)] The effect of $h_y$ is compensated by the controller in \eqref{22}.
  \item[Iii)] The effect of inaccurate measurement for $y$ can be approximated and compensated.
  \item[Iiii)] Assumption 2 is made for the approximation-based control.
\end{description}

It is of interest to know whether $\gamma_y^*=0$ holds; similar to the analysis for $\gamma^*$, it is determined by the structure of $h_y(t)$. With the additive sensor fault, some parts of $h_y(t)$ may not be handled \cite{dzhao2022}. Below, we present a design example to show the indirect stabilization following Example 2.

\emph{Example 3. Following Example 2, we consider a constant bias vector in the measurement as modeled by \eqref{21}. Rewrite \eqref{20} with respect to $y$, and it comes
\begin{displaymath}
\dot y = Ay + B\left[ {f(y) + gu + \eta (y,t) + h(y)} \right] + {h_{y,e}}(t),
\end{displaymath}
where
\begin{displaymath}
\begin{array}{l}
\begin{aligned}
{h_{y,e}}(t) =&~ {h_{y,A}}(t) + {h_{y,B}}(t),\\
{h_{y,A}}(t) =&~ - As(t),\\
{h_{y,B}}(t) = &~B[f(x) + \eta (x,t) + h(x) \\
&~- f(y) - \eta (y,t)-h(y)].
\end{aligned}
\end{array}
\end{displaymath}
Directly, the portion $h_{y,B} (t)$ can be fully compensated and the portion $h_{y,A}(t)$ has to be tolerated by the system. If $\left\| {{h_{y,A}}(t)} \right\|$ is bounded, $\left\| y \right\|$ is bounded. Specifically, following Example 2 and \eqref{28}, it can be derived that
\begin{displaymath}
\begin{array}{l}
\begin{aligned}
{{\dot V}_{c,y}} \le & - \alpha (\left\| y \right\|) - \frac{{{w_\theta }}}{2}\tilde \theta _y^T{{\tilde \theta }_y} - \frac{{{w_\lambda }}}{{2\Lambda }}\tilde \lambda _y^2 + {\varepsilon _y} + {y^T}P{h_{y,A}}(t)\\
 \le & - \alpha (\left\| y \right\|) - \frac{{{w_\theta }}}{2}\tilde \theta _y^T{{\tilde \theta }_y} - \frac{{{w_\lambda }}}{{2\Lambda }}\tilde \lambda _y^2 + {\varepsilon _y}\\
 &+ \frac{1}{2}\nu _y {\left\| y \right\|^2} + \frac{1}{2}{\nu _y ^{ - 1}}{\left\| {P{h_{y,A}}(t)} \right\|^2},
\end{aligned}
\end{array}
\end{displaymath}
where $\nu _y \in R^+$. Under the condition $\frac{1}{2}\nu_y {\left\| y \right\|^2} - \alpha (\left\| y \right\|) < 0$, the final bound of $\left\| y \right\|$ can be obtained.}

\subsection{Direct Stabilization}
As the state stabilization of $x$ is expected, we will analyze the control performance in terms of $x$ with \eqref{21} and \eqref{22} for system \eqref{1}. Different from the formulation in \eqref{24}, a controller-matching formulation for \eqref{1} is presented.

Following the fact in \eqref{3}, we have
\begin{align}\label{31}
\dot V(x) = &\frac{{\partial V(x)}}{{\partial x}}\{ f(x) + g{u_0}(x) + \eta (x,t)\nonumber\\
 &- g{u_0}(x) + h(x,t) + gu(x)\} \nonumber\\
 \le & - \alpha (\left\| x \right\|) + \frac{{\partial V(x)}}{{\partial x}}\{ {h_u}(t) + g{u_c}(y)\}
\end{align}
where ${h_u}(t) = g{u_0}(y) - g{u_0}(x) + h(x,t)$. Regarding $h_u (t)$, the ideal control effort $u_{c,u}^*(y)$ that corresponds to the non-negative bound $\gamma_u^*$ is expected to satisfy that
\begin{eqnarray}\label{32}
\frac{{\partial V(x)}}{{\partial x}}\{ {h_u}(t) - gu_{c,u}^*(y)\}  \le \gamma_u^*.
\end{eqnarray}
The optimal approximation for $u_{c,u}^*(y)$ is denoted as ${\hat u_{c,u}}(y,{\theta _u})$, which has the same structure of ${\hat u_c}(x,\vartheta )$ in \eqref{11}, and
\begin{displaymath}
{\theta _u} = \arg \left\{ {\mathop {\inf }\limits_{\vartheta  \in {\Theta }} \left\{ {\mathop {\sup }\limits_{y \in Y} \left\| {u_{c,u}^*(y) - {{\hat u}_{c,u}}(y,\vartheta )} \right\|} \right\}} \right\},
\end{displaymath}
where $\theta_u$ is the optimal weight vector for the approximation of $u_{c,u}^*(y)$. The following Assumption is made based on Assumptions 1 and 2.

\textbf{Assumption 3.} For $y \in Y$, it holds that $\left\| {u_{c,u}^*(y) - {{\hat u}_{c,u}}(y,{\theta _u})} \right\| \le {\lambda _u}$, where ${\lambda _u} \in {R^ + }$ is the unknown bounding parameter.

For the closed-loop system with \eqref{21} and \eqref{22}, consider the Lyapunov function ${V_{c,u}} = V(x) + \frac{1}{2}\tilde \theta _u^T{\Gamma ^{ - 1}}{\tilde \theta _u} + \frac{1}{{2\Lambda }}\tilde \lambda _u^2$, where ${\tilde \theta _u} = \hat \theta  - {\theta _u}$ and ${\tilde \lambda _u} = \hat \lambda  - {\lambda _u}$. Note that, now $\hat \theta $ and $\hat \lambda $ are the estimations of $\theta _u$  and  $\lambda _u$, respectively. The time derivative of $V_{c,u}$ can be re-deduced based on \eqref{31} and \eqref{32}, where
\begin{align}\label{34}
{{\dot V}_{c,u}} \le & - \alpha (\left\| x \right\|) + \frac{{\partial V(x)}}{{\partial x}}\{ {h_u}(t) + g{u_c}(y)\} \nonumber\\
 &+ {\dot {\hat \theta} ^T}{\Gamma ^{ - 1}}{{\tilde \theta }_u} + {\Lambda ^{ - 1}}{{\tilde \lambda }_u}\dot {\hat \lambda} \nonumber\\
 \le & - \alpha (\left\| x \right\|) + p(x)\{ u_{c,u}^*(y) + {u_c}(y)\} \nonumber\\
 &+ {\dot {\hat \theta} ^T}{\Gamma ^{ - 1}}{{\tilde \theta }_u} + {\Lambda ^{ - 1}}{{\tilde \lambda }_u}\dot {\hat \lambda }+\gamma_u^*\nonumber\\
 \le & - \alpha (\left\| x \right\|) + p(x)\{ u_{c,u}^*(y) - {{\hat u}_{c,u}}(y,{\theta _u})\nonumber\\
 &+ {{\hat u}_c}(y,{\theta _u}) + {u_c}(y)\}  + {\dot {\hat \theta} ^T}{\Gamma ^{ - 1}}{{\tilde \theta }_u} \nonumber\\
 &+ {\Lambda ^{ - 1}}{{\tilde \lambda }_u}\dot {\hat \lambda}+\gamma_u^*.
\end{align}
Based on Assumption 3 and $u_c (y)$, \eqref{34} can be derived as
\begin{align}\label{35}
{{\dot V}_{c,u}} \le & - \alpha (\left\| x \right\|) + \left\| {p(x)} \right\|{\lambda _u} + p(x)\nonumber\\
 &\times \{ \Pi (y)(\hat \theta  - {{\tilde \theta }_u}) - \Pi (y)\hat \theta  - \hat \lambda s(y)\}\nonumber \\
 &+ {\dot {\hat \theta} ^T}{\Gamma ^{ - 1}}{{\tilde \theta }_u} + {\Lambda ^{ - 1}}{{\tilde \lambda }_u}\dot {\hat \lambda} +\gamma_u^*\nonumber\\
 \le & - \alpha (\left\| x \right\|) + \left\| {p(x)} \right\|{\lambda _u} - \hat \lambda p(x)s(y)\nonumber\\
 &+ [p(y) - p(x)]\Pi (y)\tilde \theta  - {w_\theta }{{\hat \theta }^T}{{\tilde \theta }_u}\nonumber\\
 &- \frac{{{w_\lambda }{{\tilde \lambda }_u}\hat \lambda }}{\Lambda } + {{\tilde \lambda }_u}p(y)s(y)+\gamma_u^*.
\end{align}
Without loss of generality, assume that $\frac{{\partial V(x)}}{{\partial x}}$ is Lipschitz. Since $g$ is known, we have $p(x)$ to be Lipschitz with known (piecewise) Lipschitz constant $l_p$ and $p(0)=0$. Based on $l_p$, one has
\begin{eqnarray}\label{36}
[p(y) - p(x)]\Pi (y)\tilde \theta  \le {l_p}{\pi _m}\left\| {\tilde \theta } \right\|\left\| s \right\|,
\end{eqnarray}
and
\begin{align}\label{37}
\left\| {p(x)} \right\|&{\lambda _u} - \hat \lambda p(x)s(y) + {{\tilde \lambda }_u}p(y)s(y)\nonumber\\
&\le {\lambda _u}m\omega c + {\lambda _u}{l_p}\bar s\left\| x \right\| + \sqrt m {l_p}\left\| {{{\tilde \lambda }_u}} \right\|\left\| s \right\|,
\end{align}
where ${\pi _m} \ge \left\| {\Pi (y)} \right\|$ (bounded basis functions are used) and
\begin{displaymath}
\bar s = \left\{ \begin{array}{l}
2\sqrt m ,~~~\omega \left\| s \right\| > 2\sqrt m, \\
\omega \left\| s \right\|,~~else.
\end{array} \right.
\end{displaymath}
Based on the inequalities in \eqref{36} and \eqref{37} and completing the squares for \eqref{35}, one has
\begin{align}\label{38}
{{\dot V}_{c,u}} \le & - \alpha (\left\| x \right\|) + \frac{{{\chi _2}}}{2}{x^T}x - \frac{{{w_\theta }}}{2}\tilde \theta _u^T{{\tilde \theta }_u}\nonumber\\
 &+ \frac{{{\chi _1}}}{2}\tilde \theta _u^T{{\tilde \theta }_u} - \frac{{{w_\lambda }}}{{2\Lambda }}\tilde \lambda _u^2 + \frac{{{\chi _3}}}{2}\tilde \lambda _u^2 + {\varepsilon _u},
\end{align}
where ${\chi _1}$, ${\chi _2}$, and ${\chi _3}$ are positive scalars and
\begin{displaymath}
\begin{array}{l}
\begin{aligned}
{\varepsilon _u} = &~\frac{{{w_\theta }}}{2}\theta _u^T{\theta _u} + \frac{{{w_\lambda }}}{{2\Lambda }}\lambda _u^2 + \frac{{l_p^2\pi _m^2}}{{2{\chi _1}}}{s^T}s\\
 &+ \frac{{l_p^2\lambda _u^2}}{{2{\chi _2}}}{{\bar s}^2} + \frac{{l_p^2m}}{{2{\chi _3}}}{s^T}s + {\lambda _u}m\omega c+\gamma_u^*.
\end{aligned}
\end{array}
\end{displaymath}
Based on ${\chi _1}$, ${\chi _2}$, and ${\chi _3}$, to ensure the stability of the control system, there exists an function ${a_u}(\left\| x \right\|)$ and scalars ${w_{\theta ,\chi }}$ and ${w_{\lambda ,\chi }}$ such that
\begin{eqnarray}\label{39}
\left\{ \begin{array}{l}
{a_u}(\left\| x \right\|) = \alpha (\left\| x \right\|) - \frac{{{\chi _2}}}{2}{x^T}x>0,\\
{w_{\theta ,\chi }} = \frac{{{w_\theta }}}{2} - \frac{{{\chi _1}}}{2}>0,\\
{w_{\lambda ,\chi }} = \frac{{{w_\lambda }}}{{2\Lambda }} - \frac{{{\chi _3}}}{2}>0.
\end{array} \right.
\end{eqnarray}
Finally, taking \eqref{39} into \eqref{38} yields
\begin{eqnarray}\label{40}
{\dot V_{c,u}} \le  - {\alpha _u}(\left\| x \right\|) - {w_{\theta ,\chi }}\tilde \theta _u^2{\tilde \theta _u} - {w_{\lambda ,\chi }}\tilde \lambda _u^2 + {\varepsilon _u}.
\end{eqnarray}
If $\left\| s(t) \right\|$ is bounded and \eqref{39} holds, inequality \eqref{40} implies that system \eqref{1} with controller \eqref{22} and measurement \eqref{21} is uniformly ultimately bounded \cite{dzhao2022}. Since $\left\| s(t) \right\|$ may not be zero, the asymptotic convergence, as discussed for the accurate measurement-based control, is not possible. Or, under the fact that ${\chi _1}$, ${\chi _2}$, and ${\chi _3}$ cannot be arbitrarily large due to the stability constraints given in \eqref{39}, an arbitrary small final bound for system state is not possible, even if ${w_\theta } = {w_\lambda } = 0$ and $\omega  \to 0$. If $\alpha$ is of a quadratic form, inequalities in \eqref{39} can be solved based on the linear matrix inequality method.

For system \eqref{1} with controller \eqref{13} and measurement $y$ as in \eqref{21}, it is known that
\begin{description}
  \item[Di)] The controller in \eqref{22} can stabilize the system state $x$ and the effect of $h_u(t)$ is compensated.
  \item[Dii)] The system state $x$ as well as controller state estimation errors, ${\tilde \theta _u}$ and ${\tilde \lambda _u}$, can be uniformly ultimately bounded.
  \item[Diii)] Assumption 3, constraints in \eqref{39}, and boundedness of $s$ are needed to keep the control performance.
\end{description}

\textbf{Remark 3.} To get a deep insight into the state stabilization and most importantly, the physical interpretation and stability condition, we did not substitute $s$ from $y$ to get the performance result for $x$. Besides, in the literature, the direct stabilization analysis is mainly adopted, and this will be discussed in the following section.

\textbf{Remark 4.} For $h_u (t)$, two special cases will be discussed. If ${u_0}(y) = {u_0}(x) = 0$, the full approximation-based control is designed, where the effect of $u_0$ will be integrated into $u_c$. Thus, $h_u(t)$ is a function vector of $x$ and the ideal compensation control effort depends only on the state variable, i.e., $\frac{{\partial V(x)}}{{\partial x}}\{ {h_u}(t) - gu_{c,u}^*(x)\}  \le \gamma_u^*$. The stability analysis will follow these for the direct stabilization given above to quantify the unmatching effect of the controller due to measurement inaccuracy, and the stability conclusion, as well as the controller structure, holds the same. To be concise, this case is omitted here. If $h(x,t)=0$, it means no unknown system dynamics. However, due to the sensor fault or cyber attack (e.g., false data injection attack and deception attack), $h_u(t) \ne 0$. In this case, the compensation control effort $u_c(y)$ is aimed to \emph{revise the control signal} automatically, i.e., modifying the control signal drifting due to inaccurate measurement or any control signal implementation error (e.g., the actuator fault or the fragile controller error). Besides, it can be found that the control signal drifting can be fully covered by the proposed adaptive approximation-based control scheme. Also, since the matching condition is met, the actuator fault can be fully compensated when the measurement is accurate.

\section{Generalization Discussions}
In this section, the comparison discussions about the results given by Sections II, III.A, and III.B, based on the related literature, regarding the stability conclusion, physical explanations/interpretation, and approximation ability/assumptions, are given.

\subsection{Stability}
Based on Theorem 1, we know that the adaptive approximation-based control driven by the accurate measurement is powerful and efficient. When measurement inaccuracy occurs, without control system reconfiguration, the adaptive approximation-based controller can guarantee a certain control performance under some conditions, due to the robustness and learning ability of the controller. Specifically, the indirect stabilization analysis tells the performance bound for the adaptive approximation-based control with inaccurate measurement here, where the asymptotically tracking of the (inaccurate) measurement can be achieved automatically. The direct stabilization analysis proves the result of indirect stabilization performance analysis from the aspect of stability condition (performance limit exits due to the stability constraint), and furthermore, proves the learning ability of the approximation scheme. All in all, if a bounded inaccuracy happens to the measurement, the adaptive approximation-based control scheme can ensure a bounded control performance without controller reconfiguration.

For the inaccuracy, we would like to mention two representative cases. First, the bounded measurement noise can be tolerated by the controller; however, an unbiased filter will be helpful to further suppress the noise effect on the final bound of system state. Second, if $s(t)$ is state related error due to signal transmission delay or event-triggered transmission, the performance should be reconsidered by taking the stability margin of the controller into account \cite{tab2007,etcsurvey2018}. It should be noted that if a non-zero and non-state portion exists in $s$, e.g., the constant triggering threshold for state signal transmission, the state stabilization performance has its limit, as shown in Section III.B \cite{etcp2016} (see the explanations in the following section).
\subsection{Assumption Discussion}
From the aspect of pure functional approximation, Assumption 1 is easy to be satisfied, as the approximator will not be under actuated: the degree of parameter freedom for $h$ will not be larger than the independent measurement dimension $n$. For Assumptions 2 and 3, as the degree of parameter freedom for $h_f$ may be larger than the independent measurement dimension $n$, the pure functional approximation performance may be decreased without any control system reconfiguration. Nevertheless, it is the ideal compensation control effort that acts as the vector to be approximated (usually the control input dimension $m$ is no larger than the independent measurement dimension $n$), and the closed-loop feedback facilitates the convergence of the approximation. Besides, since the measurement space $X$ or $Y$ is considered, when $f$, $g$, $\eta$, and $h$ are local or piecewise Lipschitz, Assumptions 2 and 3 for the approximation of the optimal compensation control efforts are easy to be met.

To ensure sufficient dimensions of the inaccurate measurement, the multiplicative model is an alternative way for the inaccuracy modeling, e.g., $y = \Xi x$ ((or $u=\Xi u_0$)), where $\Xi  = diag\{ ...,{\gamma _i},...\} $, ${\gamma _i} \in [{\gamma _{i,l}},{\gamma _{i,u}}]$, ${\gamma _{i,l}}$ and ${\gamma _{i,u}}$ are nonzero lower and upper bound of ${\gamma _i}$ (either known or unknown), and $i \in \{ 1,...,n\} $. With this model, the standard adaptive control scheme can be used to get rid of uncertainty or inaccuracy resulting by $\Xi $ in both the actuator and sensor (see \cite{zhai2018,ma2018,fu2018,zhang2018,zhang22018}, and thus we can find the major literature on measurement inaccuracy and tolerant control). Note that the multiplicative model also means that no new-sourced signals are added to the measurement, and we do not need to replace the space $X$ by $Y$, even if inaccuracy occurs. Thus, sufficient measurement dimensions can be ensured. Moreover, the multiplicative model $y = \Xi x$ means a linear transformation of the state space when $\Xi$ is nonsingular (the stochastic case of $\Xi$ is out of the scope here), which does not damage/loss the state information. However, the additive inaccuracy model in \eqref{21} may lead to a state information pollution. Based on this fact, it is easy to understand why an extra-sourced inaccuracy portion in the measurement will put a limit on the control performance without control system reconfiguration.

\subsection{Unified Solution}
The tolerance and robustness of the adaptive approximation-based controller with inaccurate measurements are analyzed. It is natural to know: with the measurement model as in \eqref{21}, what will the control performance be for system \eqref{1}? Based on the literature, some related results can be classified as the direct stabilization result \cite{songsb} and the indirect stabilization result \cite{jin2017,yu2020,khe2015,khe2016,khe2018,bou2018}. Despite the difference in system description and the detailed controller parameter setting, the ``actively designed" adaptive approximation-based controller in \cite{songsb,jin2017,yu2020,khe2015,khe2016,khe2018,bou2018} corresponds to the same control performances as analyzed in this study, when inaccuracy modeled in \eqref{21} is assumed. Moreover, in this study, we point out the existence of a control performance limit for the adaptive approximation-based control systems when facing additive measurement inaccuracy. Furthermore, although the measurement inaccuracy is considered, the description of $h$, $h_y$, and $h_u$ covers the unknown modeling dynamics and control signal dynamics, e.g., process fault, disturbance, actuator fault, and cyber attack. Thus, the proposed adaptive approximation-based control scheme works for general problems.

In this case, from accurate measurement to inaccurate measurement, from direct stabilization to indirect stabilization, and from controller matching to system matching, in general the adaptive approximation-based controller owns the same controller structure and control performance, but just with different interpretations.

\textbf{Remark 5.} Additive approximation-based control effort $u_c$ is given as in \eqref{2}, which is easy to implement and restricted to no specific nominal controllers. To ensure a seamless implementation of $u_c$, it can be activated without implementation until the convergence of $u_c$.

\begin{figure*}
\begin{center}
\epsfig{file=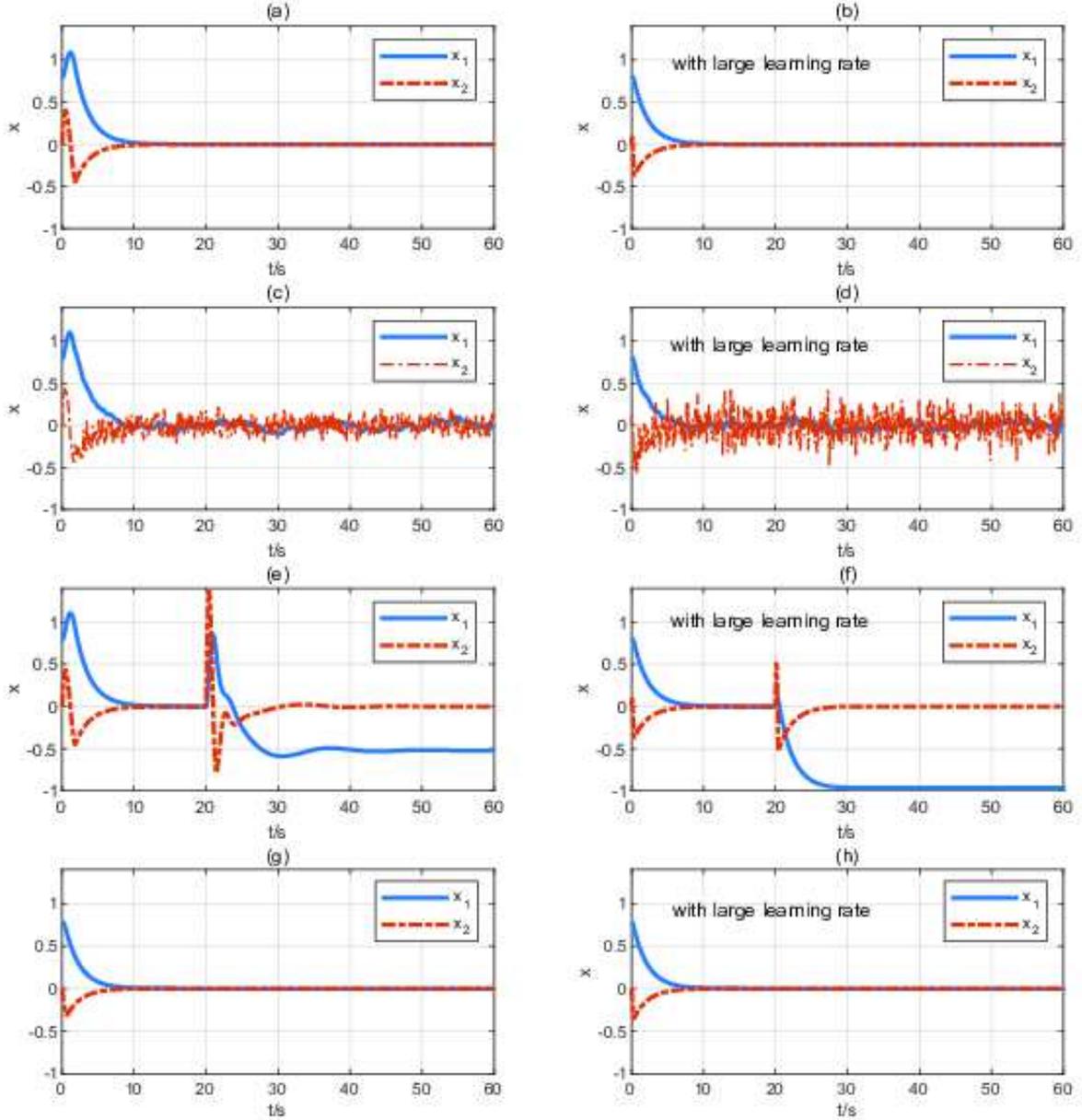,width=15.976cm,height=16.256cm}
\caption{State trajectories with different controller learning rates and measurement inaccuracies. (a) and (b): accurate measurement; (c) and (d): zero mean measurement noise with variance $0.01$; (e) and (f): sensor constant bias/disturbance/cyber attack of magnitude 1 for the measurement of $x_1$ when $t \ge 20s$; (g) and (h): fading measurement $y=0.5x$. The left column: $\Gamma=diag\{2,...,2\}$ and $\Lambda=5$; the right column: $\Gamma=diag\{20,...,20\}$ and $\Lambda=50$.}
\label{fig1}
\end{center}
\end{figure*}

\begin{figure}
\begin{center}
\epsfig{file=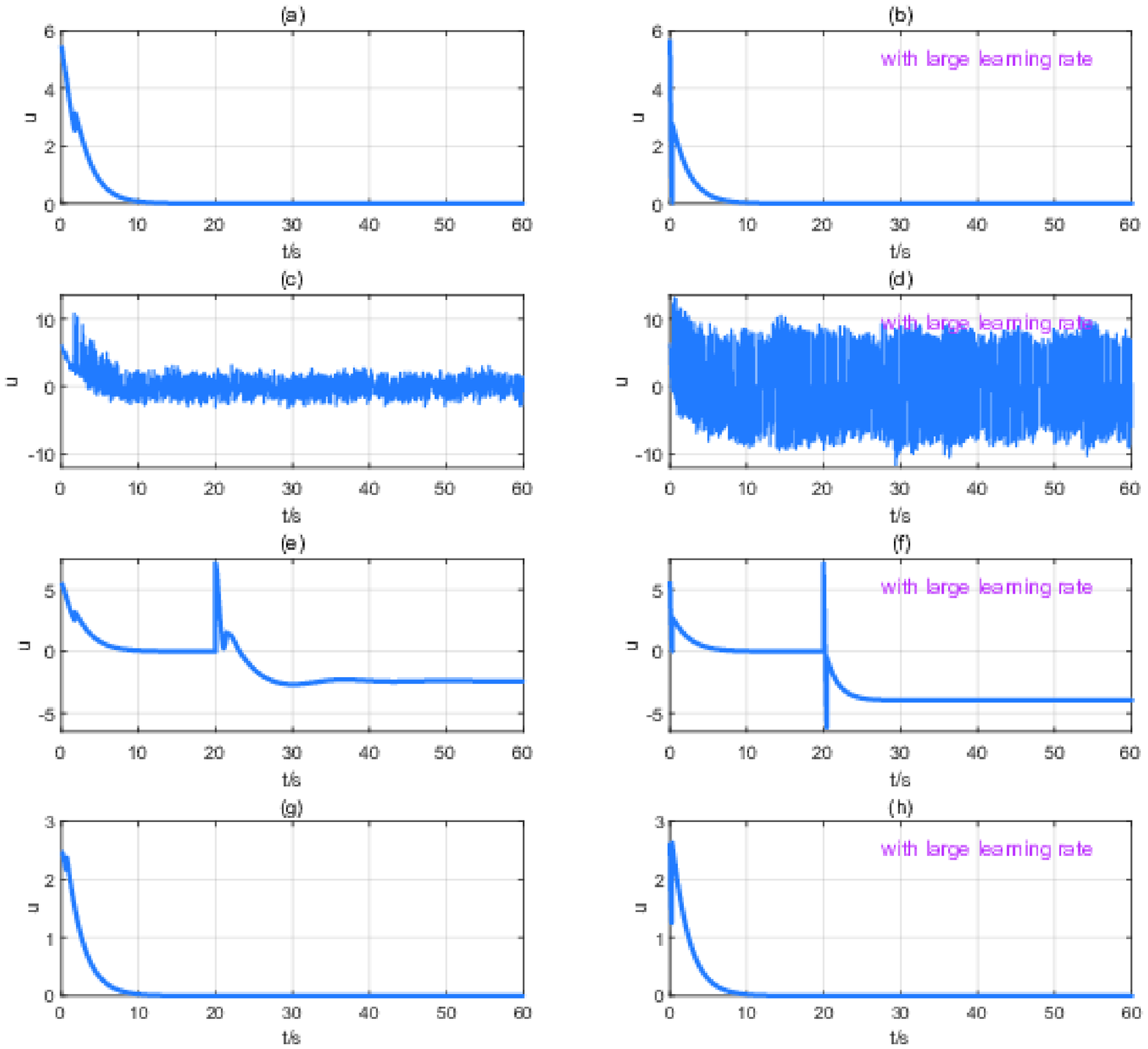,width=8.82cm,height=7.992cm}
\caption{Control input with different controller learning rates and measurement inaccuracies. (a) and (b): accurate measurement; (c) and (d): zero mean measurement noise with variance $0.01$; (e) and (f): sensor constant bias/disturbance/cyber attack of magnitude 1 for the measurement of $x_1$ when $t \ge 20s$; (g) and (h): fading measurement $y=0.5x$. The left column: $\Gamma=diag\{2,...,2\}$ and $\Lambda=5$; the right column: $\Gamma=diag\{20,...,20\}$ and $\Lambda=50$.}
\label{fig2}
\end{center}
\end{figure}

\begin{figure}
\begin{center}
\epsfig{file=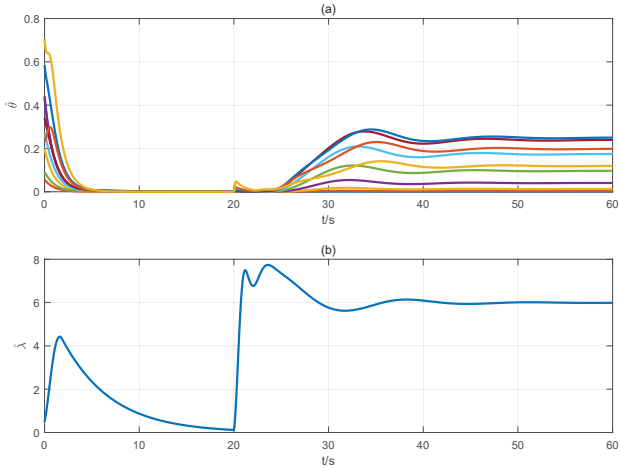,width=8.4294cm,height=6.342cm}
\caption{$\hat \theta$ (a) and $\hat \lambda$ (b). Measurement case: constant sensor bias of magnitude 1 for the measurement of $x_1$ when $t \ge 20s$; learning rates: $\Gamma=diag\{2,...,2\}$ and $\Lambda=5$.}
\label{fig3}
\end{center}
\end{figure}

\begin{figure}
\begin{center}
\epsfig{file=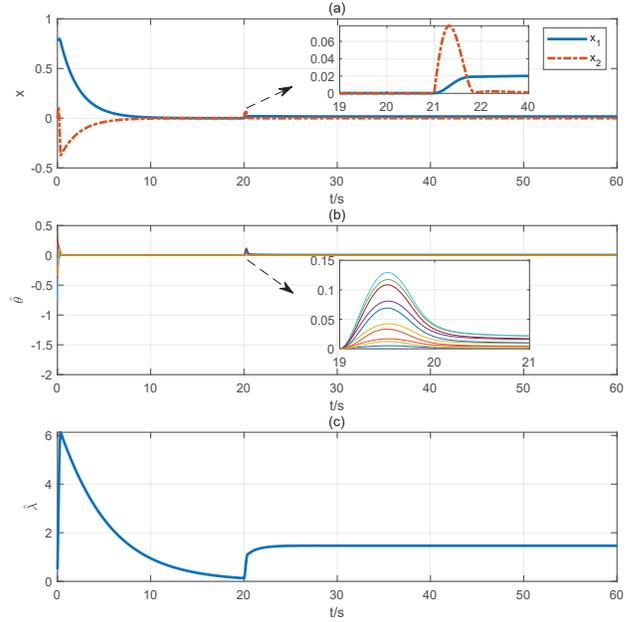,width=8.53cm,height=8.325cm}
\caption{State (a), $\hat \theta$ (b), and $\hat \lambda$ (c). Accurate measurement; $h(x,t)=5sin(x_1)+cos(x_2)+x_1^2$ when $t\ge 20s$.}
\label{fig4}
\end{center}
\end{figure}

\begin{figure}
\begin{center}
\epsfig{file=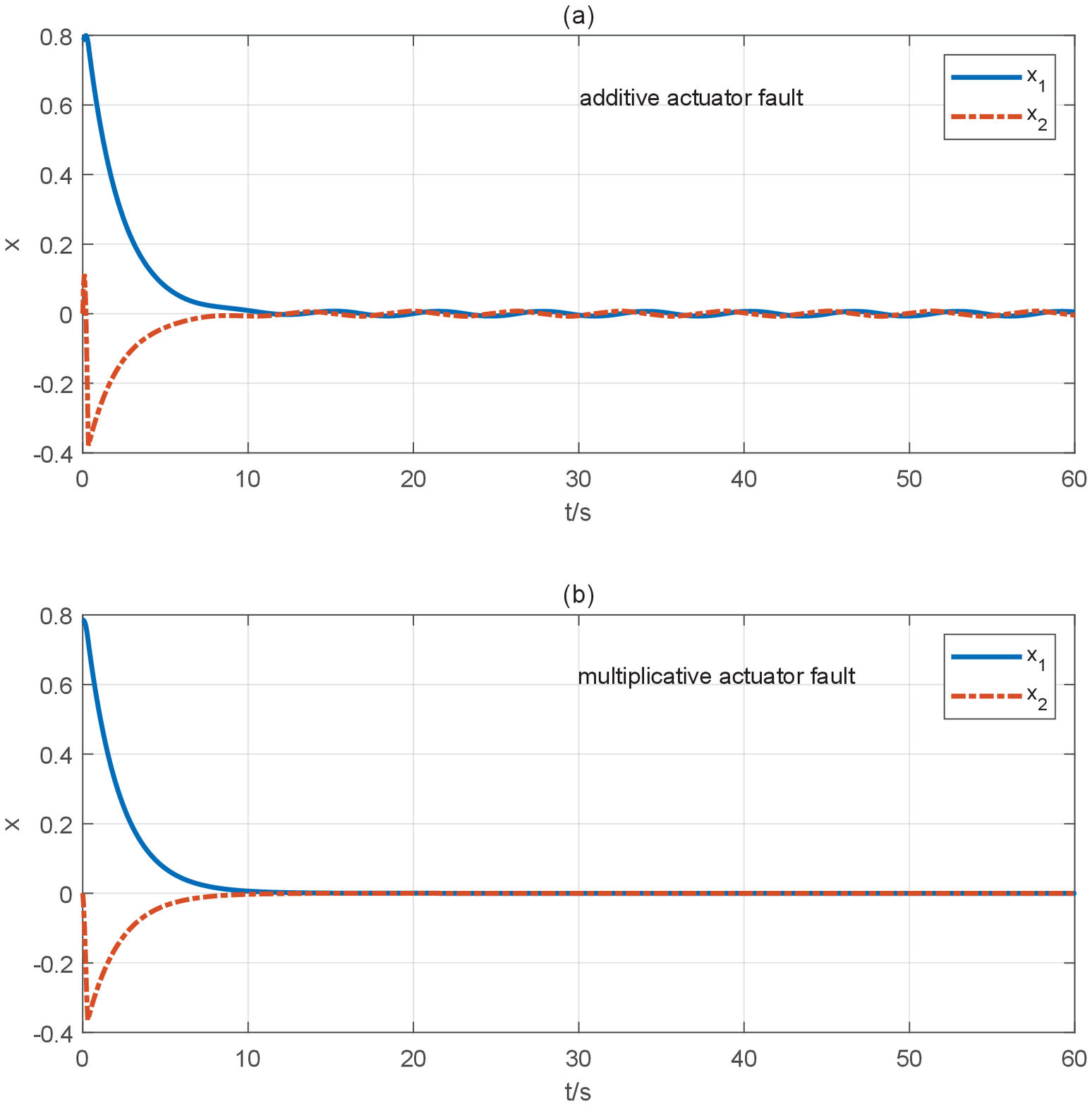,width=8.43cm,height=8.395cm}
\caption{State trajectories under actuator fault. (a) additive actuator fault $u(t)=u(t)+0.5sin(t)$; (b): multiplicative actuator fault $u(t)=0.5u(t)$.}
\label{fig4}
\end{center}
\end{figure}

\section{Simulation Study}
To verify the conclusion given in the last section, a single-link robot arm system is employed for the simulation study. The model is described by
\begin{displaymath}
\left\{ \begin{array}{l}
\begin{aligned}
{{\dot x}_1} =&~ {x_2},\\
{{\dot x}_2} =&  - \frac{{Mgl}}{J}\sin ({x_1}) - \frac{B}{J}{x_2} + \frac{{u(t)}}{J},
\end{aligned}
\end{array} \right.
\end{displaymath}
where $x_1$ is the angular position of the link, $x_2$ is the angular velocity of the link, $J$ is the total rotational inertia of the motor, $B$ is the overall damping coefficient, $M$ is the total mass of the link, $g$ is the gravitational acceleration, $l$ is the length from the joint axis to the link center of mass, and $u$ is the torque input.

For simulation purposes, the parameter setting in \cite{modelx} is adopted, where $J=1kg.m^2$, $B=2N.s/m$, $l=1m$, $M=1kg$, and $g=9.8m/s^2$. The initial value of the single-link robot system is $x_1(0)=\frac{\pi}{4}$ and $x_2(0)=0$. To check the controller performance, let the unknown system dynamics be $h(x,t)=5sin(x_1)$.

For the nominal control effort, we have $K = \left[ {\begin{array}{*{20}{c}}1&2\end{array}} \right]$ and $Q = \left[ {\begin{array}{*{20}{c}}1&1\\1&3\end{array}} \right]$ (as shown in Example 1), then we have $P = \left[ {\begin{array}{*{20}{c}}
1&{0.5}\\{0.5}&1\end{array}} \right]$. For the compensation control effort, ten Gaussian radial basis functions, uniformly centered from $-0.5$ to $0.5$ with the same width of $0.3$, are used. The controller parameters are $w_{\theta}=0.5$, $w_{\lambda}=0.2$, $\Gamma=diag\{2,...,2\}$, $\Lambda=5$, and $\omega=0.01$.

In the following, we will show the results with different measurement and process scenarios.
\subsection{Measurement Inaccuracy}
To show the effect of measurement inaccuracy on the control performance, the following cases are respectively considered: 1) accurate measurement (Figs. 1 (a) and (b)), 2) zero mean measurement noise with variance $0.01$ in both the measurement channels (Figs. 1 (c) and (d)); 3) bias sensor fault/deception cyber attack/disturbance $s(t) = {\left[ {\begin{array}{*{20}{c}}1&0\end{array}} \right]^T}$ (Figs. 1 (e) and (f)), and 4) fading measurement channel/multiplicative inaccuracy $y=0.5x$ (Figs. 1 (g) and (h)). The right column of Fig. 1 is based on the large controller learning rates $\Gamma=diag\{20,...,20\}$ and $\Lambda=50$. The corresponding control inputs are illustrated in Fig. 2. To show the learning reaction of the controller, the trajectories of $\hat \theta$ and $\hat \lambda$ corresponding to the scenario of Fig. 1 (e) are illustrated in Fig. 3.

Based on Figs. 1 (c) and (d), it is known that the adaptive approximation-based control can tolerate the zero mean measurement noise, but the tolerance decreases with the increase of the learning rate. Since a larger learning rate corresponds to better control performance, it implies a trade-off between the control performance and measurement noise tolerance. Based on Figs. 1 (e) and (f), it is clear that the system state tracks the inaccurate measurement when $t \ge 20 s$, which proves the conclusion from direct stabilization and indirect stabilization analyses. With larger learning rates, the conclusion remains the same: the state bound depends on the additive form measurement inaccuracy, where the states track the inaccurate measurements correspondingly. Figs. 1 (g) and (h) prove that the multiplicative measurement inaccuracy will not affect the convergence of the system state. Combining Figs. 1 and 2, it can be found that a larger learning rate for the controller leads to faster convergence of system state and also a more aggressive control input. Fig. 3 illustrates that the controller can re-learn and compensate for the unknown process dynamics effect in the control system due to the inaccurate measurement.

\subsection{Process and Actuator Dynamic Inaccuracy}
From Section IV, we know that unified solution/controller also covers the process dynamic inaccuracy. Based on the accurate measurement, we have $h(x,t)=5sin(x_1)$ when $t< 20s$ and $h(x,t)=5sin(x_1)+cos(x_2)+x_1^2$ when $t\ge 20s$. The state and controller state results are illustrated in Fig. 4, where bounded control performance can be ensured when a new process dynamics change occurs.

As stated in Section IV, the actuator fault is covered by the unified solution. Additive form actuator fault, $u(t)=u(t)+0.5sin(t)$, is considered and Fig. 5 (a) illustrates the result; multiplicative form actuator fault, $u(t)=0.5u(t)$, is considered and Fig. 5 (b) illustrates the result. The simulation results meet the conclusion in Section IV and verify the effectiveness of the unified solution.

\section{Conclusions}
A unified solution of adaptive approximation-based control for nonlinear uncertain systems with accurate and inaccurate measurement was obtained. It was proved that the adaptive approximation-based controller design with accurate measurement also works for the inaccurate measurement case, when certain conditions are met. The physical interpretation was given, and the control performance was quantified for this unified solution. The generalization of the unified solution was also given for actuator faults and unknown process dynamics. In the future, the observer-based adaptive approximation-based control will be further studied to reduce the conservativeness in the measurement process.

\ifCLASSOPTIONcaptionsoff
  \newpage
\fi



\bibliographystyle{IEEEtran}
\bibliography{unified_solution}
\end{document}